\documentclass[twocolumn,showpacs,preprintnumbers,superscriptaddress,amsmath,amssymb]{revtex4}

\usepackage{graphicx}

\usepackage{epsfig,latexsym,amssymb,color,bm}

\usepackage{booktabs,longtable,mathrsfs}
\usepackage{multirow}

\newcommand{\rmd}{{\rm d}}

\newcommand{\be}{\begin{equation}}
\newcommand{\ee}{\end{equation}}
\newcommand{\ba}{\begin{array}}
\newcommand{\ea}{\end{array}}
\newcommand{\bn}{\begin{eqnarray}}
\newcommand{\en}{\end{eqnarray}}
\newcommand{\bt}{\begin{tabular}}
\newcommand{\et}{\end{tabular}}
\newcommand{\bml}{\begin{mathletters}}
\newcommand{\eml}{\end{mathletters}}
\newcommand{\bc}{\begin{center}}
\newcommand{\ec}{\end{center}}
\newcommand{\bi}{\begin{itemize}}
\newcommand{\ei}{\end{itemize}}

\date{\today}

\begin{document}

\title{Effective theory for low-energy nuclear energy density functionals}

\author{J. Dobaczewski}
\affiliation{Institute of Theoretical Physics, Faculty of Physics, University of Warsaw, ul. Ho{\.z}a 69, PL-00681 Warsaw, Poland}
\affiliation{Department of Physics, PO Box 35 (YFL), FI-40014
University of Jyv{\"a}skyl{\"a}, Finland}

\author{K. Bennaceur}
\affiliation{Universit\'e de Lyon, F-69003 Lyon, France;
Institut de Physique Nucl\'eaire de Lyon, CNRS/IN2P3, Universit\'e Lyon 1,
F-69622 Villeurbanne Cedex, France}

\author{F. Raimondi}
\affiliation{Department of Physics, PO Box 35 (YFL), FI-40014
University of Jyv{\"a}skyl{\"a}, Finland}

\begin{abstract}
We introduce a new class of effective interactions to be used within
the energy-density-functional approaches. They are based on
regularized zero-range interactions and constitute a consistent
application of the effective-theory methodology to low-energy
phenomena in nuclei. They allow for defining the order of expansion
in terms of the order of derivatives acting on the finite-range
potential. Numerical calculations show a rapid convergence of the
expansion and independence of results of the regularization scale.
\end{abstract}

\pacs{21.60.Jz, 21.30.Fe}

\maketitle

A consistent formulation of the low-energy energy-density-functional
(EDF) approach in terms of effective theory is long overdue in
nuclear physics. Effective theories are now used in different domains
of physics with a tremendous success. They quantify and formalize
basic physical intuitions about aspects of the description that are
essential, and those that can be treated in terms of controlled and
correctible approximations. One spectacular example relates to
derivations of nuclear forces based on the chiral perturbation theory
augmented by the effective-field-theory treatment of corrective
terms~\cite{[Epe09],[Epe10]}. Analogous effective approach can also
be applied in a much simpler context of quantal one-body
dynamics~\cite{[Lep97]}.

Successful derivations from chiral effective field theory of
relativistic~\cite{[Fin06],[Fin07]} and
nonrelativistic~\cite{[Sto10],[Geb11]} nuclear EDFs were recently
accomplished. They lead to local EDFs with density-dependent
couplings associated with the underlying pion-exchange interaction.
These approaches strive to map out the in-medium nucleonic effects at
the two-pion-mass (2$m_\pi\simeq260$\,MeV/c$^2$) or Fermi-momentum
($k_F\simeq260$\,MeV/$\hbar{}c\simeq1.3$\,fm$^{-1}$) scales. However,
the energy scale of low-energy nuclear phenomena in finite nuclei is
much lower. For example, to dissociate a nucleon from a nucleus one
has to increase its kinetic energy by
$\delta{}E_{\text{kin}}=\hbar^2{}k_F\delta{}k/M\simeq0.25\hbar{}c\,\delta{}k$,
that is, a typical separation energy of
$\delta{}E_{\text{kin}}\simeq8$\,MeV corresponds to the momentum
increase of $\delta{}k\simeq32$\,MeV/$\hbar{}c$. Moreover, the
low-energy EDFs are supposed to describe nuclear excitations and
shell-effects at even lower energies of 1\,MeV or
$\delta{}k\simeq4$\,MeV/$\hbar{}c$ and below.

Therefore, what in the QCD-driven chiral dynamics is the small-energy
scale, becomes a short-range high-energy scale of nucleon-nucleon
force acting on relatively weakly bound nucleons in nuclei. The
question of bridging these two different energy scales is the subject
of current intense studies in nuclear physics~\cite{[Fur10]}. In
finite nuclei, surface effects decrease the infinite-matter particle
binding energies by about a factor of two. In the present study, we
aim at building a phenomenological low-energy EDFs, with gradient
terms providing for the surface terms, corresponding the nuclear
leptodermous expansion~\cite{[Gra82],[Rei06]}.

Ideas of the effective theory have recently been applied in
constructing the EDFs based on higher-order derivatives of nuclear
one-body densities~\cite{[Car08]}. When analyzed in terms of the
density-matrix expansion~\cite{[Neg72]}, the obtained quasilocal EDFs
converge fast with the order of derivatives~\cite{[Car10e]},
which is a very encouraging feature of the approach. The quasilocal
EDFs are also obtained by the Hartree-Fock averaging of higher-order
zero-range pseudopotentials~\cite{[Rai11a]}.

However, one aspect of the approach developed so far is still
missing, namely, the quasilocal EDFs and zero-range pseudopotentials
do not provide us with any expansion scale, against which the
higher-order derivatives could be classified. In the chiral effective
field theory, such a scale is given by the
lesser of the cutoff and the intrinsic breakdown scale
in the
momentum-space integrals, and allows for formulating a consistent
power-counting scheme of the diagrammatic expansion~\cite{[Wei90b]}.
In low-energy nuclear physics, an analogous cut-off regularization of
the zero-range interactions has already been
proposed~\cite{[Mog10],[Mog12]}, and the dimensional regularization
was also studied~\cite{[Mog12a]}, however, a consistent expansion
scheme is not yet available.

In the present study, we introduce the expansion scale by employing
the regularized higher-order zero-range (pseudo)potentials -- exactly
as it was illustrated in the simple examples discussed in
Ref.~\cite{[Lep97]}. Among a plethora of various ways to regularize
delta interactions, we consider the one based on the Gaussian
function,
\begin{equation}
\delta(\bm{r})
= \lim_{a\rightarrow\,0} g_a(\bm{r})
= \lim_{a\rightarrow\,0}
\frac{e^{-\frac{\bm{r}^2}{a^2}}}{\left(a\sqrt{\pi}\right)^3}  .
\end{equation}
Then, the resulting central two-body regularized pseudopotential reads,
\begin{eqnarray}
\label{eq:1}
&& \hspace*{-0.5cm} V(\bm{r}'_1,\bm{r}'_2;\bm{r}_1,\bm{r}_2) =  \\
&& \sum_{i=1}^4  \hat{P}_i
\hat{O}_{i}(\bm{k}',\bm{k})\delta(\bm{r}'_1-\bm{r}_1)\delta(\bm{r}'_2-\bm{r}_2)
g_a(\bm{r}_1-\bm{r}_2) , \nonumber
\end{eqnarray}
where
$\bm{k}=\frac{1}{2i}(\bm{\nabla}_1-\bm{\nabla}_2)$ and
$\bm{k}'=\frac{1}{2i}(\bm{\nabla}'_1-\bm{\nabla}'_2)$ are the
standard relative-momentum operators, and the Wigner,
Bartlett, Heisenberg, and Majorana terms are given by the
standard spin and isospin exchange operators,
$\hat{P}_1\equiv 1$, $\hat{P}_2\equiv\hat{P}_\sigma$,
$\hat{P}_3\equiv -\hat{P}_\tau$, $\hat{P}_4\equiv -\hat{P}_\sigma\hat{P}_\tau$.
The relative-momentum operators are the building blocks of the scalar
higher-order derivative operators $\hat{O}_{i}(\bm{k}',\bm{k})$.
Generalization of the pseudopotential to spin-orbit and tensor terms are
straightforward~\cite{[Rai11a]}, and in the present study are not
discussed.

In our definition of the pseudopotential (\ref{eq:1}), operators
$\hat{O}_{i}(\bm{k}',\bm{k})$ act on the product of the locality
delta functions
$\delta(\bm{r}'_1-\bm{r}_1)\delta(\bm{r}'_2-\bm{r}_2)$ and
regularized delta function $g_a(\bm{r}_1-\bm{r}_2)$. Derivatives of
the delta functions have to be understood in the usual sense of
distributions, namely, when the pseudopotential~(\ref{eq:1}) is
inserted into a matrix element, or when it is averaged with respect to
densities, the integration by parts transfers all derivatives onto
wave functions or densities. After the integration by parts, the
locality deltas reduce the integral to two variables,
$\bm{r}_1$ and $\bm{r}_2$, only.

To give a specific example, up to the second-order, that is,
up to the next-to-leading-order (NLO) expansion, operators
$\hat{O}_{i}(\bm{k}',\bm{k})$ read
\begin{equation}
\label{eq:6}
\hat{O}_{i}(\bm{k}',\bm{k}) = T^{(i)}_0 +\frac{1}{2}T^{(i)}_1
\left({{\bm{k}'}^*}^2+\bm{k}^2\right)+T^{(i)}_2{\bm{k}'}^*\cdot\bm{k},
\end{equation}
where $T^{(i)}_k$ are the channel-dependent coupling
constants. In the limit of $a\rightarrow0$, Eq.~(\ref{eq:6}) links the
regularized pseudopotential~(\ref{eq:1}) to the standard zero-range
Skyrme interaction~\cite{[Ben03]}.

Before considering the general pseudopotential~(\ref{eq:1}), let us
first assume that the differential operators $\hat{O}_{i}(\bm{k}',\bm{k})$
depend only on the sum of relative momenta,
$\hat{O}_{i}\left(\bm{k}+\bm{k}'\right)=\hat{O}_{i}\left(\bm{k}-{\bm{k}'}^*\right)$. At NLO,
this corresponds to second-order coupling constants obeying condition
$T^{(i)}_2=-T^{(i)}_1$. Such particular differential
operators commute with the locality deltas
$\delta(\bm{r}'_1-\bm{r}_1)\delta(\bm{r}'_2-\bm{r}_2)$, and thus can
be applied directly onto the regularized delta
$g_a(\bm{r}_1-\bm{r}_2)$. In such a case, pseudopotential~(\ref{eq:1})
reduces to a simple local potential
\begin{equation}
\label{eq:2}
V(\bm{r}) = \sum_{i=1}^4 \hat{P}_i  V_{i}(\bm{r}),
          = \sum_{i=1}^4 \hat{P}_i \hat{O}_{i}(\bm{k}) g_a(\bm{r}) ,
\end{equation}
where $V_{i}(\bm{r})=\hat{O}_{i}(\bm{k}) g_a(\bm{r})$ are
functions of the relative distance $\bm{r}=\bm{r}_1-\bm{r}_2$.
Moreover, since $\hat{O}_{i}(\bm{k})$ are scalar differential
operators, the potentials must have forms of power series of
Laplacians $\Delta$ in $\bm{r}$, that is,
\begin{equation}
\label{eq:3}
V_{i}(\bm{r}) =
\sum_{n=0}^{n_{\text{max}}} V^{(i)}_{2n} \Delta^n g_a(\bm{r}) ,
\end{equation}
where $V^{(i)}_{2n}$ are the coupling constants at order $2n$.

Within the spirit of the effective theory, the series in
Eq.~(\ref{eq:3}) is cut at a given $n_{\text{max}}$.
Then, at a given fixed scale $a$, coupling constant $V^{(i)}_{2n}$
have to be adjusted to data.
This means defining large sets of experimental observables and
optimizing values of coupling constants so as to obtain the closest
possible reproduction of the experiment, similarly as it is presently
performed for standard EDFs (see, e.g., recent Ref.~\cite{[Kor12]}).
If the expansion is relevant, the calculated observables
should only weakly depend on the regularization scale $a$, provided
it is neither too large, when the expansion scale intrudes into the
physical scale of low-energy phenomena, nor too small, where
unresolved high-energy features are incorrectly described.

Before such a complete program is carried out, in the present
exploratory study we use the fact that a successful parametrization of
the local potential already exists in the form of the Gogny
interaction~\cite{[Gog75],[Ber91b]}, whose
finite-range part can be written as
\begin{equation}
\label{eq:5}
G(\bm{r}) = \sum_{i=1}^4 \hat{P}_i  G_i(\bm{r})
          = \sum_{i=1}^4 \hat{P}_i \sum_{k=1,2} G^{(i)}_k g_{a_k}(\bm{r}) ,
\end{equation}
where Gaussians with two ranges, $a_1=0.7$\,fm and $a_2=1.2$\,fm,
have been employed. Therefore, instead of adjusting the regularized
potential to data, we first attempt to derive it directly from the
Gogny interaction. In the present study, we focus on the finite-range
part of the Gogny interaction and we keep its zero-range
spin-orbit and density-dependent terms unchanged. We then use
the results obtained for the original Gogny interaction as metadata,
wherewith we compare those obtained for the derived regularized
potentials.

At NLO ($n_{\text{max}}=1$), the eight coupling constants,
$V^{(i)}_0$ and $V^{(i)}_2$, of the local regularized potential
(\ref{eq:2}) can be easily obtained from the eight parameters of the
central Gogny force, $G^{(i)}_1$ and $G^{(i)}_2$. Indeed, at this
order, potentials (\ref{eq:3}) are combinations of a Gaussian and
Gaussian multiplied by $\bm{r}^2$, which span similar sets of
functions as those of two Gaussians with two different ranges.

Below we use a prescription that can be
applied at any order $n_{\text{max}}$, and which is based on global
characteristics of the potentials. Namely, we determine coupling
constants $V^{(i)}_{2n}$ in Eq.~(\ref{eq:3}) by requiring that the
lowest moments of both potentials are equal, that is,
\begin{equation}
\label{eq:7}
M_{2m}^{(i)} \equiv \int \bm{r}^{2m} G_i(\bm{r}) \rmd^3 \bm{r}
               =    \int \bm{r}^{2m} V_i(\bm{r}) \rmd^3 \bm{r} ,
\end{equation}
for $m=0,1,\ldots,n_{\text{max}}$. This conditions gives the
coupling constants of the regularized potential in simple
analytical forms,
\begin{eqnarray}
\label{eq:8}
V^{(i)}_{2n}
&=& \sum_{m=0}^n \left(-\frac{a^2}{4}\right)^{n-m}
\frac{M_{2m}^{(i)}}{(n-m)!(2m+1)!} \nonumber \\
&=& \frac{1}{4^n n!}\sum_{k=1,2} G^{(i)}_k\left(a^2_k-a^2\right)^n ,
\end{eqnarray}
where $G^{(i)}_k$ and $a_k$ are the parameters of the Gogny
interaction, as in Eq.~(\ref{eq:5}).

To determine properties of finite nuclei, in the present study we
used the EDF approach for regularized potential defined in
Eq.~(\ref{eq:2}), along with the coupling constants (\ref{eq:8})
derived for the D1S Gogny force~\cite{[Ber91b]}. Numerical
calculations were performed by implementing regularized potentials in
the code HFODD (v2.53r) \cite{[Sch12],[Dob12]}, which was adapted to
solving self-consistent equations for finite-range
interactions.

\begin{figure}[htb] \centering
\includegraphics[width=\columnwidth]{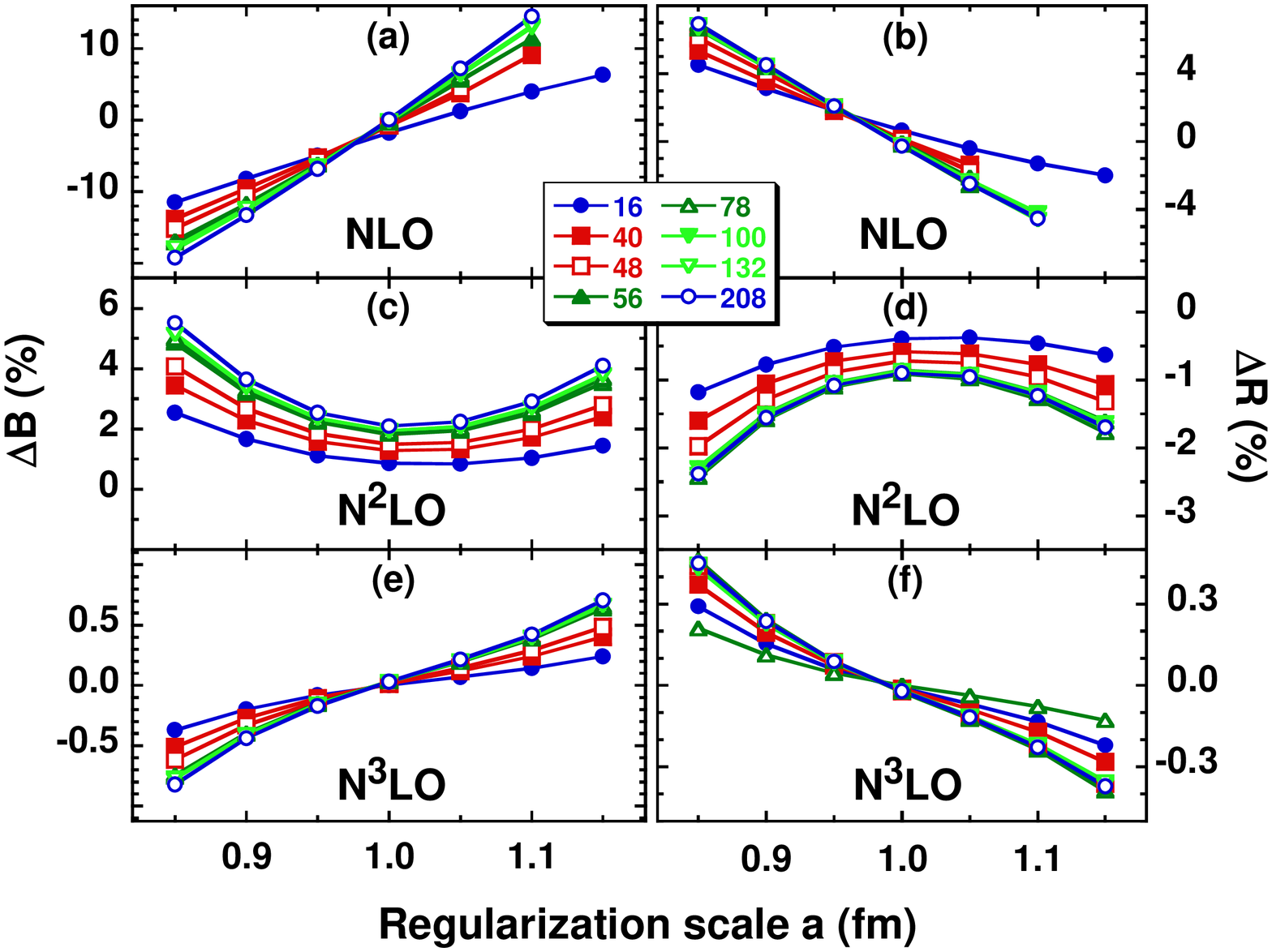}
\caption{(Color online) Deviations (in percent) of the EDF binding
energies [left panels (a), (c), and (e)] and rms radii [right panels
(b), (d), and (f)], calculated for
the regularized potential (\protect\ref{eq:2}), relative to values
obtained with the original Gogny interaction.
Top, middle, and
bottom panels correspond to the NLO, N$^2$LO, and N$^3$LO expansions,
that is, to $2n_{\text{max}}$=2, 4, and 6, respectively.
The legend gives the
atomic masses of doubly magic nuclei listed in the text. At NLO,
missing points correspond to large values of $a$, where converged
self-consistent solutions could not be obtained.
} \label{fig1}
\end{figure}

In Fig.~\ref{fig1}, we present results of calculations
performed for eight doubly magic nuclei: $^{16}$O, $^{40}$Ca,
$^{48}$Ca, $^{56}$Ni, $^{78}$Ni, $^{100}$Sn, $^{132}$Sn, and
$^{208}$Pb. In each case, we show relative deviations of binding
energies (left panels) and rms radii (right panels) as compared to
values obtained with the original Gogny interaction. One can see
that with increasing order of expansion, the convergence is very
rapid; indeed, for every next order, the obtained relative deviations
decrease by about a factor of four, and at N$^3$LO they are well
below 1\%.

We note here that the results presented in Fig.~\ref{fig1} were
obtained without any adjustment of parameters. However, one can expect
that such small adjustments can easily correct for
the smooth trends observed in Fig.~\ref{fig1}. This is so because
binding energies of nuclei, which result form cancellations of the
kinetic and potential energies, are extremely sensitive to parameters
of the interaction. For example, we checked that modifications of the
Gogny density-dependent term of the order of 0.05\% are able to bring
the N$^3$LO mass of $^{208}$Pb exactly to the Gogny value.

\begin{figure}[htb] \centering
\includegraphics[width=\columnwidth]{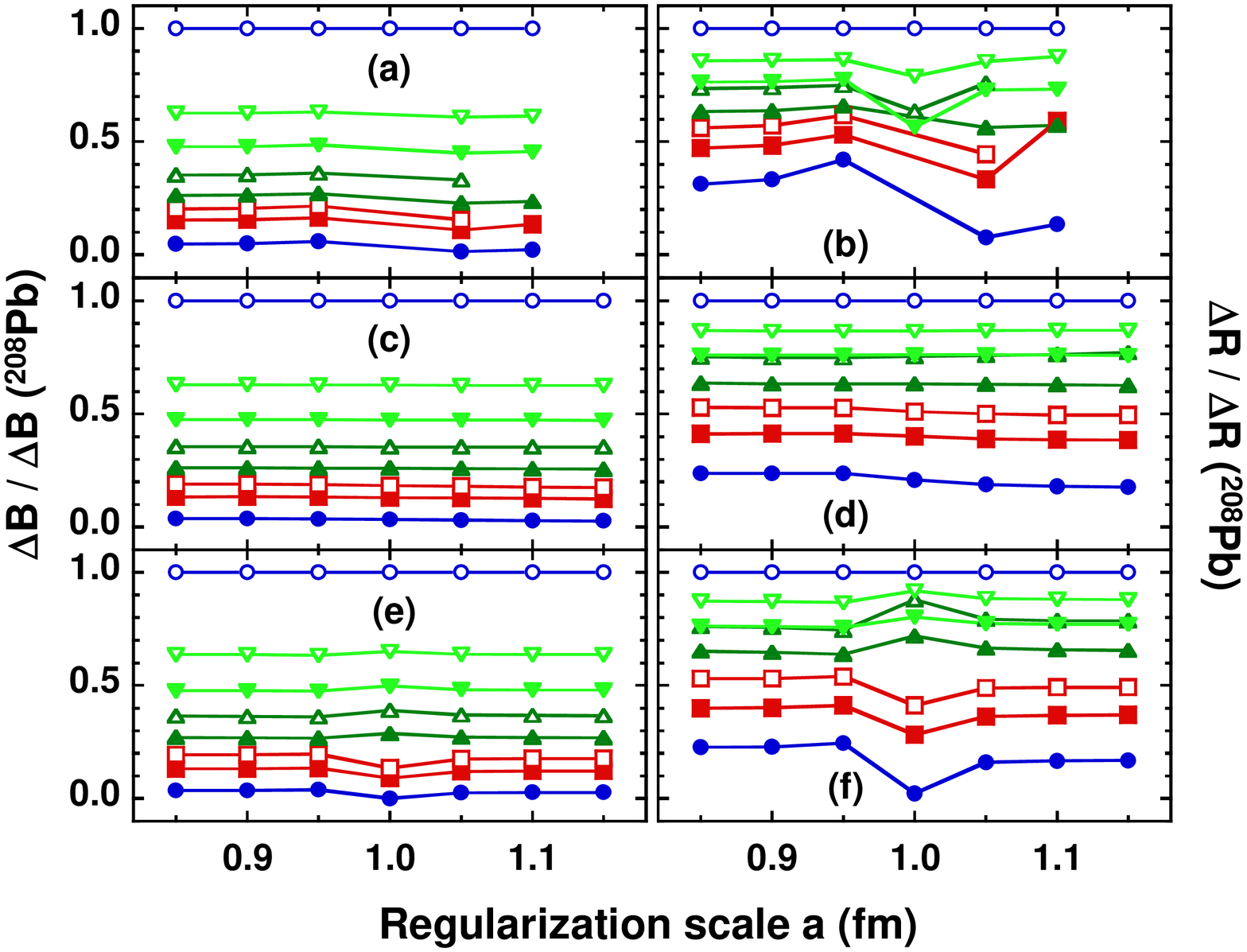}
        \caption{(Color online) Same as in Fig.~\protect\ref{fig1},
but for the deviations relative to those obtained for $^{208}$Pb.
For clarity, near $a\simeq 1$\,fm, where the $^{208}$Pb normalization values
cross zero (cf.~Fig.~\protect\ref{fig1}), several points were removed
from the plot.
} \label{fig3}
\end{figure}

The regularized potentials proposed in this study (with or without
the restricted dependence on the relative momenta introduced
above), will later be used in extensive optimization procedures --
similarly as it has been done for other standard functionals used
within the EDF approaches. Nevertheless, even without such extensive
adjustments, basic features of the obtained effective theory can
already be illustrated by studying scaling properties of the results.
To this end, in Fig.~\ref{fig3} we show deviations of binding
energies and radii relative to those obtained for $^{208}$Pb. One can
see that the results scaled in this way are almost independent of the
regularization scale, which is an extremely encouraging result,
consistent with the effective-theory principles. We also see that the
flatness of lines shown in Fig.~\ref{fig3} allows for picking
any nucleus as a reference -- so our choice of $^{208}$Pb is, in
fact, irrelevant.

Effective theories are built to provide us with expansions that
converge better when the bound-state energy or scattering energy is lower. Spectacular
examples of such a convergence pattern exist for weakly bound electron states
in quantum mechanics of a one-body Coulomb problem~\cite{[Lep97]} or
for phase shifts of the two-body nucleon-nucleon
scattering~\cite{[Epe10]}. Convergence of an effective theory can be
best visualized by the so-called Lepage plot~\cite{[Lep97]}, which
shows the dependence of the error in the description of a given
observable on energy. Such error should decrease with decreasing
energy and with increasing order of expansion.

In many physical systems, scale of energy is inverted with
respect to the scale of distance, that is, small distances correspond
to large energies and {\it vice versa}. In finite nuclei, the situation is
more complex, because the average energy per particle and average
internucleon distance are almost constant throughout the mass chart,
which is a simple consequence of the saturation of nuclear forces and
near-constancy of density inside nuclei. Should then the effective
theories work better in light (small) nuclei or in heavy (large)
ones?

Good arguments in favor of the second option are provided by the
density-matrix expansion (DME)~\cite{[Neg72]}, which uses the fact
that the range of nuclear forces is smaller than characteristic
distances at which densities in nuclei vary, see
Refs.~\cite{[Dob10],[Car10e]} and references cited therein, and by the ideas
of the leptodermous expansion~\cite{[Gra82],[Rei06]}. Indeed, since the
largest variations of densities occur at nuclear surfaces, effective
theories should work best in large nuclei, where the surface region
is relatively small with respect to the nuclear bulk.

\begin{figure}[htb] \centering
\includegraphics[width=\columnwidth]{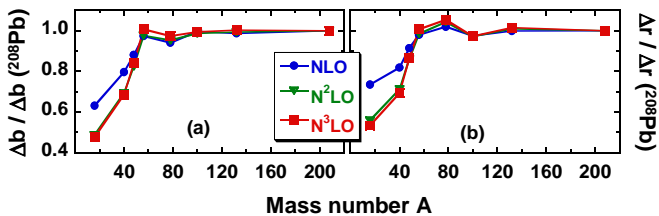}
        \caption{(Color online) Same as in Fig.~\protect\ref{fig3},
but for the deviations of binding energies scaled by the numbers
of particles, $\Delta b=\Delta B/A$, and those of radii
scaled by the nuclear sizes, $\Delta r=\Delta R/A^{1/3}$. Note
that the N$^2$LO results almost perfectly hide behind those obtained
at N$^3$LO.
} \label{fig4}
\end{figure}

To study this
problem, in Fig.~\ref{fig4} we show reduced deviations of binding
energies, $\Delta b=\Delta B/A$, and radii, $\Delta r=\Delta
R/A^{1/3}$, in function of the particle number $A$. This allows us to
compare results obtained for different nuclei in the same scale.
Since the relative deviations are almost independent of the
regularization scale $a$ (cf.~Fig.~\protect\ref{fig3}), here we
plot them at $a=0.85$\,fm.

We see that the results obtained for nuclei
beyond about $A=48$ scale with mass as those for $^{208}$Pb. From the
point of view of the effective theory and proposed expansion, one can
interpret this result as a high similarity of these nuclei, that is,
their sizes are not different enough to induce significant
differences in their convergence properties, and they all would
appear at the same point of the Lepage plot. We also see that
properties of lighter nuclei, with $A\leq 48$, converge slightly
better than those of the heavier ones. However, only a future complete
analysis, performed for the entire mass table and for optimized
functionals, can here provide conclusive results.

It is worth recalling at this point that the EDFs are designed to be
used within the variational approach and that the variation over
one-body densities leads to mean-field-like equations. Nevertheless,
the obtained total energies and one-body densities are supposed to be
exact, at least in principle, provided a hypothetical exact EDF is
used. Therefore, the regularized pseudopotential introduced in this
study should not be confused with the real two-body (effective)
interaction, for which the many-body perturbation theory (MBPT) is in
order, and for which the mean field only gives the first-order
approximation. This means that some technology and methodology
developed in standard effective theories, e.g., corrections based on
diagrammatic expansions, may not directly apply here.

The effective-theory expansion introduced in this study pertains to
resolving finer and finer details of one-body density matrices,
whereas the regularization scale $a$ pertains to folding these
details away. Thus large $a$ would correspond to too coarse a
smearing of one-body densities (that is, intrude in low-energy
dynamics), and small $a$ would too much accentuate variations thereof
(that is, introduce unphysical high-energy effects). We see that the
inapplicability of the MBPT methods does not hamper the use of
effective-theory ideas in the EDF theory.

We complete our analysis by discussing values of coupling constants
(\ref{eq:8}) in the so-called natural units, see
Refs.~\cite{[Fur97a],[Kor10c]} and references cited therein.
Since the EDF generated by our regularized potential contains
only terms quadratic in densities, in the natural units
the coupling constants have the following dimensionless values,
\begin{equation}
\label{eq:9}
v^{(i)}_{2n}= f^2 \Lambda^{2n} V^{(i)}_{2n},
\end{equation}
where $f^2$ is an overall normalization factor and $\Lambda^2$ is
a characteristic scale of the derivative operator
$\Delta$.

\begin{figure}[htb] \centering
\includegraphics[width=0.7\columnwidth]{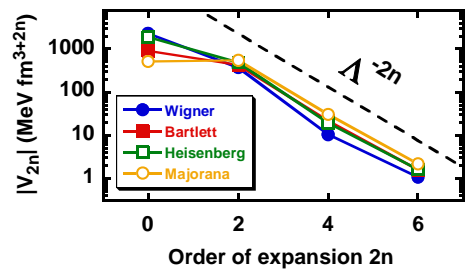}
\caption{(Color online) Coupling constants $V^{(i)}_{2n}$
derived at $a=0.85$\,fm from the Gogny interaction, as dictated by
Eq.~(\protect\ref{eq:8}). The dashed line illustrates the slope of
the $n$-dependence of the coupling constants, plotted here for
$\Lambda\simeq700$\,MeV/$\hbar{}c$.
} \label{fig5}
\end{figure}

Coupling constants $V^{(i)}_{2n}$ (\ref{eq:8}) are plotted in
Fig.~\ref{fig5}.
In the logarithmic scale they decrease almost linearly with $n$.
A rough adjustment of the slope of this decrease gives
$\Lambda^{-2n}$ for $\Lambda\simeq700$\,MeV/$\hbar{}c\simeq3.5$\,fm$^{-1}$.
For our regularized potentials at N$^3$LO, we thus obtain a similar scale
$\Lambda$ as that characterizing the zero-range interactions at NLO
($2n_{\text{max}}$=2)~\cite{[Fur97a],[Kor10c]}. Next, we fix the
normalization factor $f$ so as to make the values of dimensionless
coupling constants $v^{(i)}_{2n}$ of the order of unity, which gives
$f\simeq35$\,MeV/$(\hbar{}c)^{3/2}\simeq1$/(77\,MeV$^{1/2}$fm$^{3/2}$).
As we see, this value is significantly smaller than the pion decay
constant $f_\pi\simeq93$\,MeV/$(\hbar{}c)^{3/2}$, which normalizes
the EDFs derived directly from the chiral effective field
theory~\cite{[Fur97a]}.

\begin{figure}[htb] \centering
\includegraphics[width=\columnwidth]{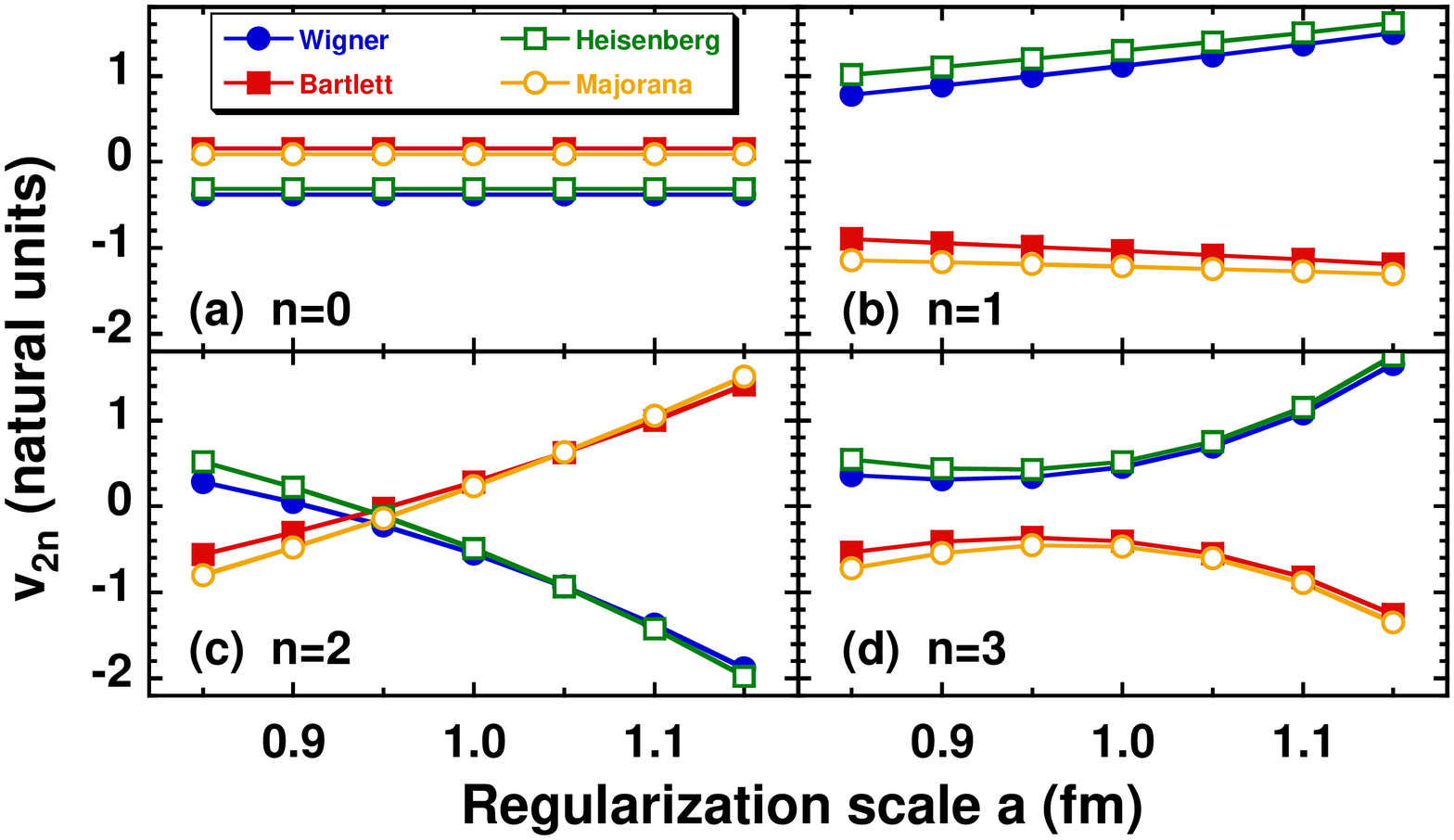}
\caption{(Color online) Coupling constants $v^{(i)}_{2n}$ in natural
units (\protect\ref{eq:9}) as functions of the regularization scale
$a$, plotted for $n=0$--3 in panels (a)--(d).
} \label{fig6}
\end{figure}

Finally, for the derived values of $\Lambda$ and $f$, in
Fig.~\ref{fig6} we show the dependence of the dimensionless coupling
constants $v^{(i)}_{2n}$ on the regularization scale $a$. We see that
for $n=1$, 2, and 3 they have values of the order of unity at all scales,
although the
fourth-order coupling constants ($n=2$) cross zero around
$a=0.95$\,fm.
The zero-order coupling constants ($n=0$) appear to be less natural,
with values of the order of 0.1 (cf.~also Fig.~\ref{fig5}).
However, we should keep in
mind that in the present study we derived the coupling constants from
the Gogny interaction, which has a predefined scale of the order of
1\,fm. Therefore, we can expect that the future adjustments of the
coupling constants directly to data may lead to even weaker
scale-dependence.

In summary, in this study we introduced a new class of energy
density functionals that are based on pseudopotentials related to
regularized delta interactions. We used the regularization scheme
employing the standard Gaussian form, which introduces a
regularization scale $a$ corresponding to the range of the
interaction. Different orders of expansion then correspond to
different orders of derivatives used in the pseudopotential.

In our opinion, future prospects for using the proposed regularized
(pseudo)potentials are high. First, similarly as in the simple
one-body examples~\cite{[Lep97]}, they may present better convergence
properties than similar expansions based on the zero-range
interactions. Second, they allow for convergent summations of
contributions from high single-particle momenta, which is not the
case for zero-range interactions. And third, they allow for
formulating a consistent expansion in terms of the orders of
derivatives, with the convergence properties gauged against the
regularization scale. This last feature is unique among all the EDF
approaches based on zero-range and finite-range interactions
developed so far, and gives us a potential of building an
order-by-order correctible theory.

Interesting comments by Dick Furnstahl and Witek Nazarewicz are gratefully acknowledged.
This work has been supported in part by the Academy of Finland and
University of Jyv\"askyl\"a within the FIDIPRO programme, and by
the Polish French agreement COPIN-IN2P3 Project No.\ 11-143. We acknowledge the CSC
- IT Center for Science Ltd, Finland for the allocation of
computational resources.

\bibliographystyle{unsrt}

\end{document}